# Precipitating Ordered Skyrmion Lattices from Helical Spaghetti

Dustin A. Gilbert,[1,2] [*] Alexander J. Grutter,[1] Paul M. Neves,[3] Guo-Jiun Shu,[4,5] Gergely Zimanyi,[6]

Brian B. Maranville,[1] Fang-Cheng Chou,[4] Kathryn Krycka,[1] Nicholas P. Butch,[1,3]

Sunxiang Huang,[6] Julie A. Borchers[1]

[1] *NIST Center for Neutron Research, National Institute of Standards and Technology, Gaithersburg, MD, 20899*
[2] *Department of Materials Science and Engineering, University of Tennessee, Knoxville, TN*
[3] *Center for Nanophysics and Advanced Materials, Department of Physics, University of Maryland, MD*
[3] *Center for Condensed Matter Sciences, National Taiwan University, Taipei, 10617 Taiwan*
[4] *Department of Materials and Mineral Resources Engineering, Institute of Mineral Resources Engineering, National Taipei University of Technology, Taipei 10608, Taiwan*
[5] *Department of Physics, University of California, Davis, CA*
[6] *Department of Physics, University of Miami, FL*

*dagilbert@utk.edu

Subject Areas: Condensed Matter Physics - Magnetism

**Abstract**

Magnetic skyrmions have been the focus of intense research due to their potential applications in ultra-high density data and logic technologies, as well as for the unique physics arising from their antisymmetric exchange term and topological protections. In this work we prepare a chiral jammed state in chemically disordered (Fe, Co)Si consisting of a combination of randomly-oriented magnetic helices, labyrinth domains, rotationally disordered skyrmion lattices and/or isolated skyrmions. Using small angle neutron scattering, (SANS) we demonstrate a symmetry-breaking magnetic field sequence which disentangles the jammed state, resulting in an ordered, oriented skyrmion lattice. The same field sequence was performed on a sample of powdered $Cu_2OSeO_3$ and again yields an ordered, oriented skyrmion lattice, despite relatively non-interacting nature of the grains. Micromagnetic simulations confirm the promotion of a preferred skyrmion lattice orientation after field treatment, independent of the initial configuration, suggesting this effect may be universally applicable. Energetics extracted from the simulations suggest that approaching a magnetic hard axis causes the moments to diverge away from the magnetic field, increasing the Dzyaloshinskii-Moriya energy, followed subsequently by a lattice re-orientation. The ability to facilitate an emergent ordered magnetic lattice with long-range orientation in a variety of materials despite overwhelming internal disorder enables the study of skyrmions even in imperfect powdered or polycrystalline systems and greatly improves the ability to rapidly screen candidate skyrmion materials.

**Introduction**

Magnetic skyrmions have emerged as a promising foundation for ultra-low power next-generation memory and logic devices[1-10] and have attracted significant interest for fundamental research due to their exotic antisymmetric exchange interaction and nontrivial topology.[11-15] In natural skyrmion materials, highly-ordered monodomain states are prepared by entering the skyrmion stability envelope in temperature and magnetic field, with the field oriented orthogonal to a crystallographic plane with low symmetry in the magnetocrystalline anisotropy.[16] Choosing a low-symmetry orientation nucleates skyrmion domains with a common preferential lattice orientation, promoting a monodomain configuration and demonstrating the critical role of the magnetocrystalline coupling in these materials. The skyrmion lattices can be subsequently re-oriented within this plane by introducing a symmetry-breaking anisotropy using an electric field[17], uniaxial pressure,[18] or spin-transfer torque.[19] These approaches require that the material have limited internal disorder and sufficient magnetocrystalline anisotropy to define the orientation of the skyrmion lattice. For materials with weak anisotropy or high internal disorder,[7,20-22] the symmetry may be broken in many directions throughout the sample, resulting in a multi-domain state. It has been suggested that this multidomain state consists of rotationally-disordered skyrmion lattices,[23] but it may include other chiral phases depending on the boundary between domains and on the preparation sequence.

The B20-structured intermetallic compound (Fe, Co)Si is unique in that it is a skyrmion material with significant internal disorder even in crystalline form. In this material the Fe and Co atoms randomly occupy the *A*-site position in the B20 structure, resulting in a random distribution of exchange (defined by the exchange stiffness, $A$) and Dzyaloshinskii-Moriya interactions (DMI, defined by the DMI coefficient, $D$), and magnetocrystalline anisotropy ($K_U$). Due to the short-range nature of these interactions (typically approximated by the exchange length, $l \approx 5$ nm) compared to the periodicity of the spin-texture (typically >50 nm), the physical characteristics of the chiral skyrmion and helix states are determined by the average of these terms. However, the local variations in *A*, *D*, and $K_U$ may act as nucleating and pinning defect sites. As typically occurs in polydomain systems, once nucleated, a magnetic domain may propagate until it intersects a pinning defect site or another domain. For misoriented helical states, including the skyrmion state, the intersection of two domains may leave neither able to propagate or reorient, becoming trapped in a jammed state, analogous to magnetic spin ice or frustrated magnets.[24,25] As a result (Fe, Co)Si can possess a complex magnetic state comprised of skyrmions, helicies,[26,27], and/or labyrinth domains (Fig. S1c), all of which possess a regular periodicity determined by the average *A* and *D*, but not a long-range orientation. Disentangling this complex state requires consideration of the role of topology,[9,12] as well as the mechanism for disentanglement, which may involve magnetic monopoles.

In this work, we demonstrate the precipitation of ordered, oriented skyrmion lattices from an otherwise disordered chiral jammed state prepared in a single-crystal of (Fe, Co)Si. Rotating the sample in a static magnetic field promotes the formation of hexagonally-ordered skyrmion lattices with a common orientation, independent of the initial orientation of the magnetic field relative to the crystal axes. Motivated by the ability to promote ordered skyrmion lattices with an orientation defined by the rotation, a similar approach was applied to a powdered sample of $Cu_2OSeO_3$. In the powdered sample the weak magnetic coupling between grains yields a state in which each grain nominally consists of a single skyrmion domain that is not aligned with its neighbors. Performing the rotational sequence on the powder also precipitated ordered and oriented skyrmion lattices despite the overwhelming structural disorder. The unexpected promotion of a preferred skyrmion orientation is verified using micromagnetic simulations. Plotting the energy terms from the simulations reveals a close relationship between skyrmion lattice orientation and the magnetic hard axis, although reorientation is promoted even without magnetocrystalline anisotropy. The dissolution of the trapped chiral domains and subsequent formation of ordered skyrmion lattices increases the overall topological charge of the system which may be accompanied by the nucleation of magnetic monopoles.

A 1.5 g single crystal ingot of $Fe_{0.85}Co_{0.15}Si$ was grown by the Bridgman method in a floating zone furnace, following previously reported methods.[28,29] The [100] and [001] axes of the crystal were identified using X-ray diffraction with Cu-Kα (1.5418 Å wavelength) radiation. Small angle neutron scattering (SANS) was performed at the NIST Center for Neutron Research on the 30 m NG7SANS beamline[30] using unpolarized 8 Å neutrons and a detector distance of 5 m. The sample was zero-field cooled to 17.5 K, then a saturating field of 500 mT was applied and then reduced to 40 mT for measurement, placing the sample within the skyrmion stability envelope.[31] The magnetic field was applied with an electromagnet along the neutron flight path and initially aligned with the crystalline [100], [110], or [111] axes, as identified in the text; additional data measured with the magnetic field orthogonal to the neutron path are shown in Supplemental Materials. Initial alignment of the sample was achieved by rotating the sample and magnet through ±1° (limited by the bore of the magnet). Over this angular range the intensity of the primary helical reflections did not change substantially, consistent with the large mosaicity values (> 5°) reported by other authors.[23] Based upon this rocking curve characterization, all measurements were made at an angular position near the rocking curve maximum that preserves right-left symmetry across the detector. We note that the right and left scattering features were determined to individually be maximal at rocking angles of ±0.63°, which are within the $q_z$ resolution of the SANS instrument when the sample is centered at a rocking angle of zero.

Except where noted, all SANS measurements were performed with the neutron beam, indicated by $n^0$, parallel to the magnetic field and orthogonal to the axis of rotation, as shown in **Figure 1a**. During the measurements, the sample was rotated 90° step-wise in the static magnetic field about the vertical $q_y$ direction. The rotation direction was then reversed and the crystal returned back to the original orientation, as shown in Fig. 1a. For example, for the sample initially aligned with $H$//[100] the rotation in the (001) plane passed through the [110] crystalline axis, ending at [010] at 90°; for the sample aligned initially with $H$//[111], the rotation in the $(1\bar{1}0)$ plane passed through the [001] and [110] axes. The SANS patterns are identified by $\phi$, the sample's cumulative angle of rotation, (sequential rotations of 90° clockwise, then 90° counter-clockwise, for example, is $\phi$=180°). The scattered intensity was plotted as a function of wavevector $q$ (where $q_y$//[001] or [1$\bar{1}$0]) or the azimuthal angle in the $q_x$-$q_y$ scattering plane, $\theta$, measured counter-clockwise relative to the $+q_x$ direction (Fig. 1a). Overall, rocking curve and orthogonal measurements (Supplemental Materials) described above revealed that the transverse width (i.e., mosaicity) of the resultant scattering features did not change substantially during rotation (i.e., with increasing $\phi$), Rotations at lower temperatures and fields are described in the Supplemental Material.

A second sample of powdered $Cu_2OSeO_3$ was fabricated from CuO and $SeO_2$ precursors as described previously.[32] The 50 mg sample comprised of misoriented crystallites with diameters of >500 nm. The powder was pressed and sintered at 573 K to form a weakly bound pellet and placed in an aluminum foil packet for measurement. SANS measurements were performed using 7 Å neutrons and a detector distance of 15 m after zero-field cooling to 57 K, applying a saturating field of 500 mT, then measuring in a field of 23 mT.[33] Due to the disk-like shape of the pressed powder, the sample is always measured in the same orientation relative to the applied field and $\phi$ is only measured in increments of 180°, which corresponds to a clockwise rotation 90° in the applied field followed by a counterclockwise rotation of 90° to return the sample to its original orientation. SANS results are processed by subtracting the scattering pattern measured above the Curie temperature from the measured data to remove structural scattering.

Simulations were performed on the object-oriented micromagnetic framework (OOMMF) platform[34-37] using a 3 nm cubic mesh, with a grain modeled as a cylinder with a 300 nm diameter and 150 nm long and a second simulation of a rectangular cuboid (parallelepiped) grain 300 nm × 300 nm × 150 nm. The surfaces along the short axis are coupled by periodic boundary conditions, effectively forming a long rod. The magnetic field is applied orthogonal to the short axis such that the skyrmions always have a finite length. For the cylinder the sample geometry makes the geometric contributions the same at every angle of rotation. A cubic magnetocrystalline anisotropy was included with a random site-by-site strength between $K_U$=0 and 50×10³ J/m³, reflecting the random Fe and Co occupancy, with the easy-axes oriented along the initial field direction and axis of rotation. We use an exchange stiffness of $1\times10^{-11}$ J/m, saturation magnetization of $1\times10^{6}$ A/m, Dzyaloshinskii-Moriya coefficient of $5\times10^{-3}$ J/m$^2$, and the applied magnetic field was 800 mT. The orientation of the resultant skyrmion lattice was determined using a fast Fourier transform (FFT). The skyrmion lattice is shown in Figure 6 from the *x* ($\phi$=0°-45°, $\phi$=135°-180°) or *y* axis ($\phi$=56°-124°), to minimize parallax.

**Results**

Magnetometry measurements of the (Fe, Co)Si crystal performed at 10 K show minimal hysteresis or remanent magnetization indicating weak magnetocrystaline anisotropy, consistent with previous results.[23] Consequently, we expect minimal coupling between the magnetic lattice and the crystalline axes.

The skyrmion state was prepared in the (Fe, Co)Si sample by zero-field cooling to 17.5 K, applying a saturating field of 500 mT parallel to the [100] axis, and then reducing the field to 40 mT. At this temperature and magnetic field the (Fe, Co)Si is expected to form skyrmion spin textures, but not skymion lattices with long-range order, shown explicitly in Fig. 1D of Ref. [31]. The magnetic field defines the orientation of the skyrmion tubes while the orientation of the skyrmion lattice is typically determined by the magnetocrystalline coupling to the crystal lattice. However, in this system, the weak magnetocrystalline anisotropy leads to poor relative alignment of the skyrmion lattices. It has been previously reported that the observed disorder is suggestive of a metastable state consistent with "weakly stratified skyrmion lines."[31,38] It is possible that the domain nucleation/propagation sequence leads to trapped regions, or domains, bounded by other misaligned domains, which possess chiral structures that are not traditional six-fold skyrmion lattices.

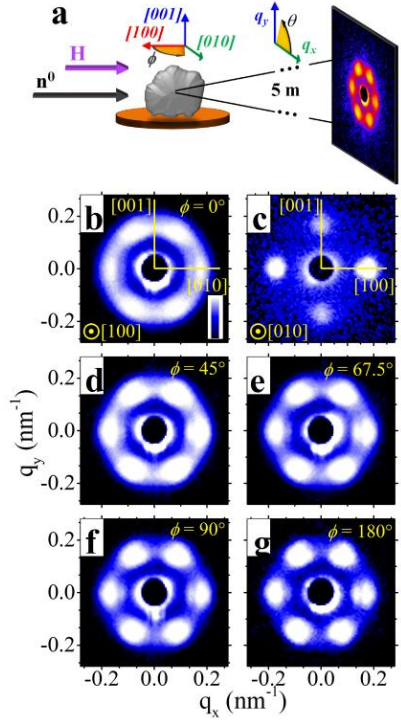

**Figure 1.** (a) Diagram of SANS measurement setup. SANS patterns for (Fe, Co)Si initially aligned with the $H$//[100] axis measured (b) with $n^0$// [100] and (c) $n^0$//[010]. SANS patterns with $H$//$n^0$ rotating the sample in the (001) plane by $\phi=$ (d) 45°, (e) 67.5°, (f) 90°, then (g) back to the initial orientation ($\phi=180°$).

SANS measurements taken in these conditions show a weak hexagonal pattern sitting atop a uniform ring of scattering, Figure 1b. The hexagonal SANS pattern indicates a hexagonally-ordered scattering potential in the plane orthogonal to the neutron beam. For this material, the hexagonally-ordered pattern has been associated with ordered skyrmion lattices with a common orientation, while the uniform ring is associated with a collection of features with no net orientation, but a regular periodicity, such as labyrinth domains of uniform width (Fig. S1c), randomly-oriented helices, or rotationally disordered skyrmion lattices (Fig. S1g). For reference, labyrinth domains have been observed in similar materials previously, and produce a ring in the Fourier transform[12], as detailed in Supplemental Materials. Disordered helicies and skyrmion lattices with a uniform periodicity can also give rise to a ring in the SANS pattern in the absence of a long-range orientation, e.g. the system must be multidomain. Multidomain skyrmions with exactly two orientations have been previously observed and yield a 12-fold SANS pattern. Attempting to fit our data (Fig. 1b) to a 12-fold pattern very poorly describes the data and results in a large $\chi^2$ value of 56, we conclude that the multidomain state must consist of more disorder than can be described by a simple two domain picture, an interpretation which is further supported by the broad azimuthal width of the six-fold peaks.

The six-fold pattern in these measurements is oriented with maxima at $\theta=0°$, $60°$ and $120°$. Since the SANS pattern is a Fourier transform of the real space structure, this orientation of the six-fold pattern corresponds to a hexagonal skyrmion lattice with nearest neighbors aligned along $\theta = 30°$, $\theta = 90°$ (the axis of rotation), and $\theta = 150°$. Details of the candidate magnetic structures and their scattering patterns are provided in the Supplemental Materials. Using simulations to construct a multidomain state consisting of labyrinth domains and rotationally disordered skyrmion lattices qualitatively reproduces the experimental observations, as demonstrated in Supplemental Figure S1.

SANS patterns measured in the orthogonal configuration (e.g. with $H//[100]$ and neutron beam along [010], Figure 1c) show two diffraction peaks along the horizontal $q_x$-axis (now corresponding to the [100]) and two along the vertical $q_y$-axis (corresponding to the [001]). The two dots along the vertical axis arise from the intersection of the scattering plane with the ring observed in Figure 1b, while the horizontal dots indicate a standard helix structure along the [100] direction. These SANS patterns together identify the presence of three magnetic features: (1) an ordered, oriented skyrmion lattice in the (100) plane, (2) a multidomain state comprised of some undetermined combination of labyrinth domains, randomly oriented helices, or rotationally disordered skyrmion lattices also in the (100) plane and (3) a helix propagating along the [100] direction. These orthogonal measurements also demonstrate that the SANS features, and associated disorder, are restricted predominantly to the plane orthogonal to the field direction, as the narrow $q_x$ width of the reflections indicates the skyrmion tubes are well aligned with the field.

Returning to the original measurement configuration described in Figure 1a and shown in Figure 1b, the sample was incrementally rotated in the (001) plane from its original alignment with $H//[100]$ to $H//[010]$, shown in Figure 1d-f. As the sample is rotated, the six-fold symmetry becomes much stronger, suggesting growth of the skyrmion phase and an enhanced collective orientation of the skyrmion domains. Rotating the sample back to $H$//[100] direction (corresponding to $\phi = 180°$), the six-fold structure gains even more intensity and becomes more pronounced, Figure 1g. At $\phi = 180°$ the sample is in its original orientation, but the scattering pattern is clearly different from that of the initially-prepared state. The observed changes thus are not an artifact of the sample geometry, but rather originate from an irreversible growth of the skyrmion phase and enhancement of its orientation. Throughout this rotation there is little change observed in measurements taken from the orthogonal perspective (in a configuration analogous to Figure 1c, with the neutron beam perpendicular to the magnetic field), shown in Figure S2 in Supporting Materials.

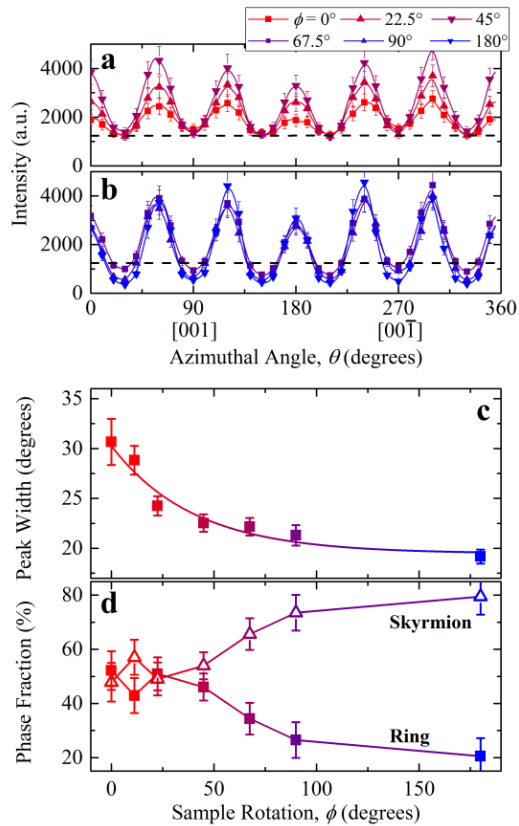

**Figure 2.** Annular averages for (a) $\phi = 0°$-45° and (b) $\phi = 67.5°$-180°. (c) The azimuthal width of the six peaks of the skyrmion phase and (d) phase fractions from the skyrmion phase and ring phase, determined from their fractional contributions to the scattering intensity.

To better identify the details of the ordering we focus on the emergence of six-fold periodicity by performing an angular-resolved azimuthal projection of the SANS patterns for $0.11$ nm$^{-1}$ $< q < 0.25$ nm$^{-1}$.

The azimuthal projection, **Figures 2a** and 2b, shows the integrated intensity in the defined $q$-range versus the angle $\theta$ in the $q_x$-$q_y$ scattering plane. In the as-prepared state the projection shows a periodic pattern well offset from the baseline. As the sample is rotated through $0° \leq \phi \leq 45°$ the amplitude of the periodic features increases while the baseline from the ring remains nearly constant, Figure 2a. It is striking that at this stage changes in the six-fold structure, which is attributable to the skyrmions, occur without corresponding changes in the ring feature, suggesting that there is no direct conversion between the magnetic phases. This behavior supports the conclusion that the ring is not due to simply rotationally disordered skyrmion lattices. Rotating the sample further, for $45° < \phi \leq 90°$ the amplitude of the oscillations remains largely constant while the ring intensity decreases, Figure 2b. Rotating the sample back to its original alignment ($\phi=180°$) the amplitude increases and the baseline decreases.

The azimuthal projection is modeled by six coupled Gaussians from the commonly-oriented skyrmions, constrained to be separated by 60°, and an angle-independent vertical offset from the ring-feature: $I(\theta) = C + \sum_{j=0}^{5} A_j\, e^{\frac{(\theta - \mu - (60j))^2}{2\sigma_j^2}}$, where $C$ corresponds to the underlying ring, $A_j$ and $\sigma_j$ are the amplitude and width of each Gaussian, respectively, and $\mu$ is the average orientation of the skyrmion lattices relative to $q_x$ axis. This fit converged for each dataset with a reduced $\chi^2$ of <1.2. The average width of the fitted Gaussians, Figure 2c, decreases by >30% during the 180° $\phi$ rotation, indicating improved rotational alignment of the skyrmion domains; the decrease is well fitted by an exponential decay function, $\sigma(\phi) = \sigma_0 + A e^{\phi/t}$, asymptotically approaching a width of 11°. Note that the fit for an alternative model with 12-peaks, representing two discrete skyrmion domains oriented along the magnetic easy axes, was extremely poor and converged to a $\chi^2$ of 56.

A simple phase fraction of commonly-oriented skyrmions versus labyrinth domains and randomly-oriented helices can be calculated by comparing the integrated weight of the individual contributions to the total scattering pattern, shown in Figure 2d. This integration posits that the mosaicity, corresponding to the distribution of the skyrmion tube alignment, remains approximately constant upon rotation, and does not include any ferromagnetic phases or helices oriented along the field direction. As described previously, this assumption is supported by both rocking curve and orthogonal measurements (Fig. S2) of the transverse width.[23] Interestingly, as the sample is rotated through the first 45° the phase fraction of oriented skyrmions changes very little, from 50% to 54%. This change is surprisingly small considering the apparent amplitude change in Figure 2a, but arises due to the commensurate decrease in the azimuthal peak width. On the other hand, the phase fraction of oriented skyrmions grows considerably over the subsequent rotation $45° < \phi \leq 180°$, reaching a maximum of 80%.

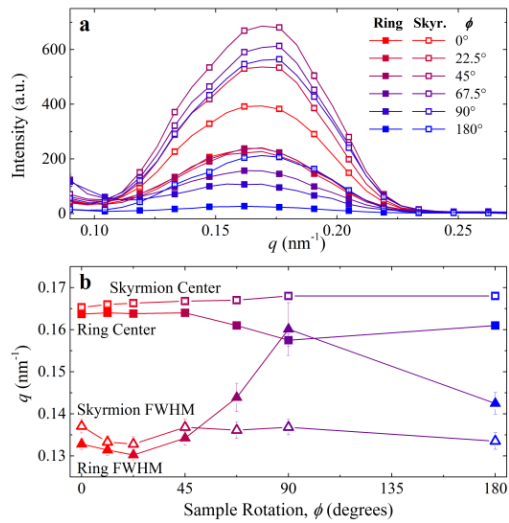

**Figure 3.** (a) Summations of radial sector averages from Figure 1b and d-g, taken at increments of 60° with $\Delta\theta=\pm 3.5°$ starting at 0° and 30°, capturing the radial distribution of the ring and skyrmion phases, respectively. Half of the data symbols are not shown for clarity. (b) Peak center and width for the sector averages of the ring and skyrmion features as a function of the sample rotation angle.

Ordering of the skyrmion features can also be investigated by averaging the intensity versus $q$ over discrete sectors. Two sector averages are presented here, overlapping with the six-fold feature and the interstitial spaces, thus capturing the skyrmion and ring features separately. Each individual sector spans a range of $\pm 3.5°$: $I(q) = \sum_{j=0}^{5} \int_{-3.5}^{3.5} I(q, 60°j + \theta) d\theta$ for the skyrmions and $I(q) = \sum_{j=0}^{5} \int_{-3.5}^{3.5} I(q, 60°j + 30° + \theta) d\theta$ for the ring. The sector averages from Figure 1b and d-g, shown in **Figure 3a**, and their collated trends with sample rotation angle $\phi$, Figure 3b, show that the ring and skyrmion features are both consistently centered at 0.164 nm$^{-1}$ ± 0.005 nm$^{-1}$, corresponding to a real-space periodicity of 77 nm for the skyrmion size and helical pitch, consistent with the previously reported skyrmion spacing for (Fe, Co)Si.[31,39] The width of the skyrmion features along the radial direction is relatively insensitive to rotation in $\phi$, indicating that this parameter is dominated by local material properties, as suggested above. Together, the trends highlighted in the azimuthal projections (Fig. 2) and sector averages (Fig. 3) imply that $\phi$–rotation causes the skyrmion domains to develop a collective orientation, and that new skyrmions are being created while the periodicity throughout the system remains relatively constant. All these observations are consistent with deconstructing a jammed state of chiral domains. Interestingly, approaching a $\phi$ of *90°* the FWHM of the ring feature increases, potentially identifying a dispersal of the chiral domain periodicity upon approaching the magnetic easy axis; we note that there is no FWHM variation around the magnetic hard axis.[27] The role of the magnetic easy and hard axes in determining the magnetic structure will be also investigated in the micromagnetic simulations.

For the investigations described above, the field was initially aligned along the high-symmetry [100] axis, which coincides with the magnetic easy axis, and the sample was rotated back-and-forth through the low-symmetry [110] direction, a local magnetic hard axis. For comparison, measurements were also performed with the magnetic field initially along the [110] direction, shown in **Figure 4**. In this configuration the plane orthogonal to the magnetic field has only one easy axis (namely the [001] direction), which would typically define the orientation of the skyrmion lattice. However, the SANS measurements (Figure 4a) once-more show both the ring and six-fold features, indicating mixed helix/labyrinth phases and rotationally disordered skyrmion lattices. As the sample is rotated by $\phi$=180° in the (001) plane, the six-fold lattice is resolved once more (Figure 4b-d), shown in detail in the azimuthal projection (Figure 4e). Rotating the sample in a static field again promotes a common orientation of the skyrmions (Figure 4f) and breaks up the disordered phase to precipitate the ordered skyrmion phase. Consequently, the skyrmion phase fraction is increased from 37% to 79%, Figure 4g. It is also noteworthy that, as in the previous case, the ring feature does not significantly decrease in magnitude during the first 45° rotation (Figs. 1 and 2). This behavior suggests that 45° of net rotation is required to significantly change the magnetic structure, regardless of the magnetocrystalline anisotropy.

Similar rotations were also performed in the (110) plane, starting with the field parallel to the [111] axis, e.g. the global magnetic hard axis, Fig. 4h. As the sample was rotated by $\phi$=90° (through the [001] axis, Fig. 4i) and $\phi$=180° (through the [110] axis, to the [$\bar{1}\bar{1}\bar{1}$] magnetic hard axis, Fig. 4j), the ring transformed to develop the six-fold pattern. After the rotation, the orientation of the skyrmion lattice is the same as that obtained after rotating in the (001) plane. The skyrmion lattice orientation in this material thus appears to be determined by the rotation direction and not by the magnetocrystalline anisotropy. In other systems such as MnSi, which possess stronger magnetocrystalline anisotropy, the crystalline axes clearly determine the skyrmion lattice orientation, even under rotation.[13,40] Outside of the skyrmion stability window, rotation of the sample promotes an oriented helix, destroying any precursor state, as shown in Figure S5.

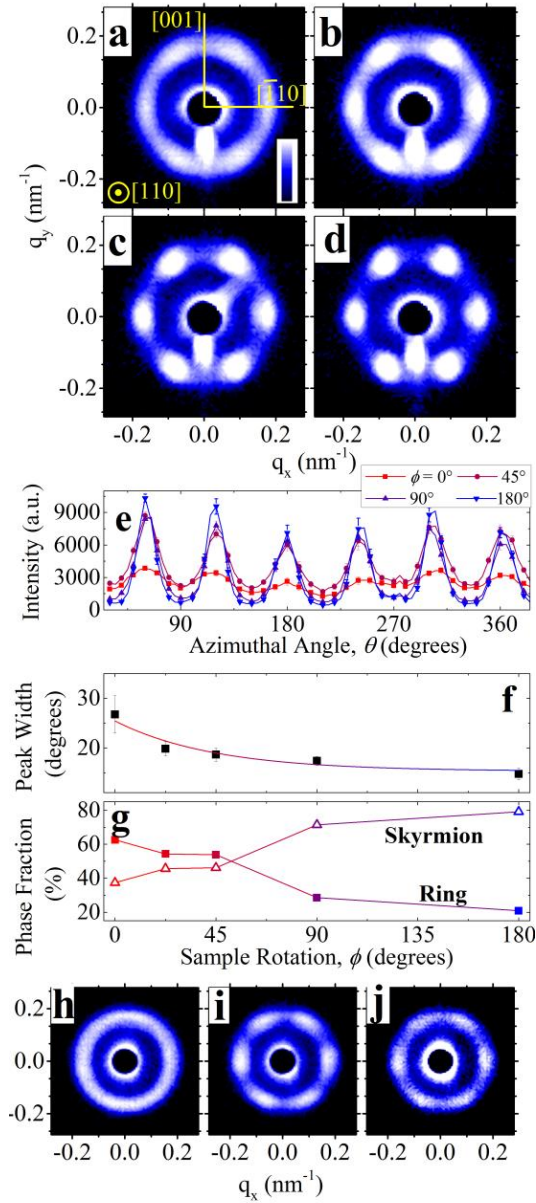

**Figure 4.** SANS patterns for (Fe, Co)Si initially aligned with the H//[110] axis measured (a) along the [110] axis, then rotated in the (001) plane by (b) 45°, (c) 90°, then (d) rotated back to 0° ($\phi$=180°). (e) The azimuthal projection of the data, (f) width of the six peaks of the skyrmion phase and (g) phase fractions from the skyrmion phase and ring phase, determined from their fractional contributions to the scattering intensity. (h) SANS pattern after saturation along the [111] axis, and rotation by (i) $\phi$=90° and (j) $\phi$=180°

Rotation-induced promotion of the skyrmion phase along both high and low-symmetry axes suggests that the skyrmion-ordering mechanism is independent of the initial orientation of the magnetic field relative to the magnetocrystalline easy/hard axes. We examine this premise in the limit of extreme disorder by performing the skyrmion-ordering sequence on a powdered sample of $Cu_2OSeO_3$. The nature of this sample - comprised of mostly segregated, randomly oriented grains - mandates that skyrmion lattices will form in the plane orthogonal to the magnetic field with random orientations and that the interactions between domains will be comparably small. To clarify, we expect each grain to consist of a single ordered skyrmion lattice, with an orientation defined by the local crystalline orientation and little to no correlation in orientation among neighboring grains; this is not equivalent to the chiral jammed state discussed above. Furthermore, since SANS captures the contributions from the whole sample, the magnetocrystalline anisotropy will not play a net role but will manifest as a highly-localized potential landscape promoting or resisting the skyrmion lattice migration during rotation. In the powder sample, symmetry breaking is realized only by the rotation in the plane orthogonal to the vertical direction in Figure 1a.

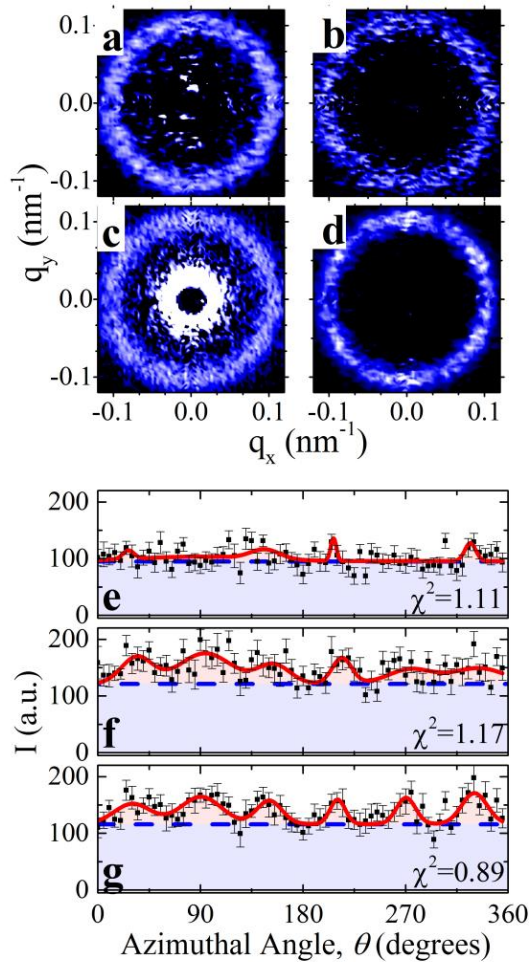

**Figure 5.** SANS patterns from powdered $Cu_2OSeO_3$ taken at (a) $\phi=0°$, (b) 360°, (c) 720°, and (d) 1440°. Azimuthal projections of the (e) $\phi=0°$, (f) 720° and (g) 1440°, obtained at $q=0.11\pm0.035$ nm$^{-1}$.

SANS measurements of the as-prepared $Cu_2OSeO_3$ sample at 57 K after zero-field cooling then applying a saturating 0.3 T magnetic field then reducing to a measurement field of 23 mT (**Figure 5a**) show the ring structure expected for a powder sample, with a radius of 0.103 nm$^{-1}$ (corresponding to a periodicity of 61 nm). Rotating the sample back and forth in increments of $\phi=90°$, through up-to 8 cycles ($\phi=1440°$, Figure 5b-d), causes the emergence and strengthening of a six-fold hexagonal pattern, indicating a net orientation of the skyrmion domains. The six-fold pattern shows maxima along $\theta=30°$, 90° and 150°, a 30° rotation relative to the (Fe, Co)Si single crystal. The Fourier transform of the SANS pattern indicates that the real space skyrmion lattice orientation has nearest neighbors aligned along the direction orthogonal to the rotation axis, e.g. in the rotation plane, along $\theta=0°$ and $\pm60°$. An azimuthal projection of the data at 0.07 nm$^{-1} \leq q \leq 0.13$ nm$^{-1}$, Figure 5e, shows no significant periodic feature in the as-prepared sample indicating that any helices and/or skyrmion domains are randomly aligned, with their orientation defined locally within each grain. Performing the rotational sequence, Figures 5f and 5g the azimuthal projection of the rotated sample has a clear six-fold periodicity; calculating the phase fraction from the feature weight, the ordered skyrmion phase with this orientation accounts for 16±0.7% of the sample.

**Discussion**

In two different systems we have experimentally demonstrated the ability to precipitate and grow ordered, oriented skyrmion lattices from disordered, co-planar chiral structures, even in the face of overwhelming intrinsic disorder, by rotating the sample in a magnetic field. In the single crystal (Fe, Co)Si the internal chemical disorder and weak magnetocrystalline coupling is expected to nucleate helical phases throughout the sample. These magnetic configurations may locally propagate but become jammed by intersecting other chiral configurations. In our experimental procedure, the sample is saturated, then the magnetic field is reduced to the nominal center of the skyrmion stability envelope. As the field is reduced, the sample will initially relax into a conical phase oriented along the field direction. Approaching the skyrmion stability envelope, helices nucleate in the plane orthogonal to the magnetic field, presumably from randomly distributed defect sites. These newly nucleated states will consist of three co-planar helices oriented at 60°, forming a skyrmion state. In the absence of a significant magnetocrystalline anisotropy there will be relatively weak relative orientation between domains of these skyrmion domains. Furthermore, the randomly distributed nucleating sites will mean that the spacing between domains will not be an integer multiple of the skyrmion pitch. Accordingly, the resultant state may possess many misaligned skyrmion domains, as well as other stable helical configurations resulting from the unmatched termination at the boundary of the domains.[31,41] The resultant magnetic configuration would become jammed, with neighboring regions possessing different magnetic configurations, unable to expand and thus lacking long-range rotational order. Rotating the samples in a magnetic field not only promotes the formation of new skyrmions, and therefore varying the total topological charge of the system, but also overcomes any jamming and causes disordered helical domains to align while conserving their topology.

Recent works have similarly observed the augmentation of skyrmion lattices in single crystals under rotation.[13,40] In these works, it was reported that the SANS pattern for MnSi undergoes a six-fold to 12-fold to six-fold sequential transformation when passing through low and high-symmetry axes. We report similar results on a MnSi single crystal in the Supplemental Material. This behavior is distinctly different than that observed for (Fe, Co)Si, which shows only a persistent decreases in the second minority phase and indifference to the crystalline axes. The contrast between the response of MnSi (Supplemental Materials and Refs. [13] and [40]) and (Fe, Co)Si to field rotation emphasizes the role of the magnetocrystalline anisotropy in skyrmion formation.

Similarly, the powdered $Cu_2OSeO_3$ sample shows an enhancement of the six-fold symmetry that emerges with rotation. Previous works, however, have shown that the rotation of a $Cu_2OSeO_3$ single crystal can cause additional peaks to emerge in the SANS pattern,[13] suggesting that the magnetocrystalline anisotropy plays a role in this system, as in MnSi. In the powdered sample, the skyrmion lattices are expected to nucleate within each grain with an orientation reflecting the local crystallographic orientation. However, the skyrmion lattice orientation is expected to be localized due to poor intergranular coupling and the energy cost in propagating lattices between neighboring grains with incommensurate orientations. One way in which the rotation and magentocrystalline anisotropy may combine to promote a net order is by breaking the symmetry of the cubic structure within each grain, thus lifting the degeneracy of equivalent easy axes via field rotation. The net result for a sum of many grains would be the depopulation of skyrmion lattices coupled to crystalline axes in the plane of rotation and the subsequent growth of those aligned towards the rotation axis. Comparing the results for (Fe, Co)Si, MnSi and $Cu_2OSeO_3$ supports recent reports emphasizing the role of anisotropy in skyrmion materials.[31,42-47]

To provide a better understanding of the experimental results, analogous simulations were performed on a cylindrical grain in a rotating magnetic field using the OOMMF simulation platform. Bloch-type skyrmions were defined in the initial conditions as downward domains, oriented along -$x$, in a matrix oriented in +$x$ with a smoothly-varying domain boundary. A cubic anisotropy was used with the magnetic easy axes along the x, y and z coordinate axes. Hexagonal lattices of skyrmions with two distinct orientations were generated (1) with the nearest neighbors adjacent along the axis of rotation at $\theta$ ($t$=0) = 30°, 90° and 150° (left column in **Figure 6a**) and (2) 30° rotated, with nearest neighbors adjacent in the plane of rotation at $\theta$ ($t$=0) = 0°, 60° and 120° (center column in Figure 6a). The magnetic configuration viewed from the axis of rotation is shown in the right column of Figure 6a, taken from (2). The skyrmion configuration in panel $i$ is the relaxed lattice resulting from the initial conditions. First considering the rotation of the skyrmion tubes (right column), the simulations show that the skyrmions tube direction, which is initially parallel to the easy [100] direction, follows the rotation of the magnetic field, consistent with the SANS results. The skyrmion cross-sections appear as ovals rather than stripes due to a deflection of the skyrmion tubes in -$z$ which manifests in the model.

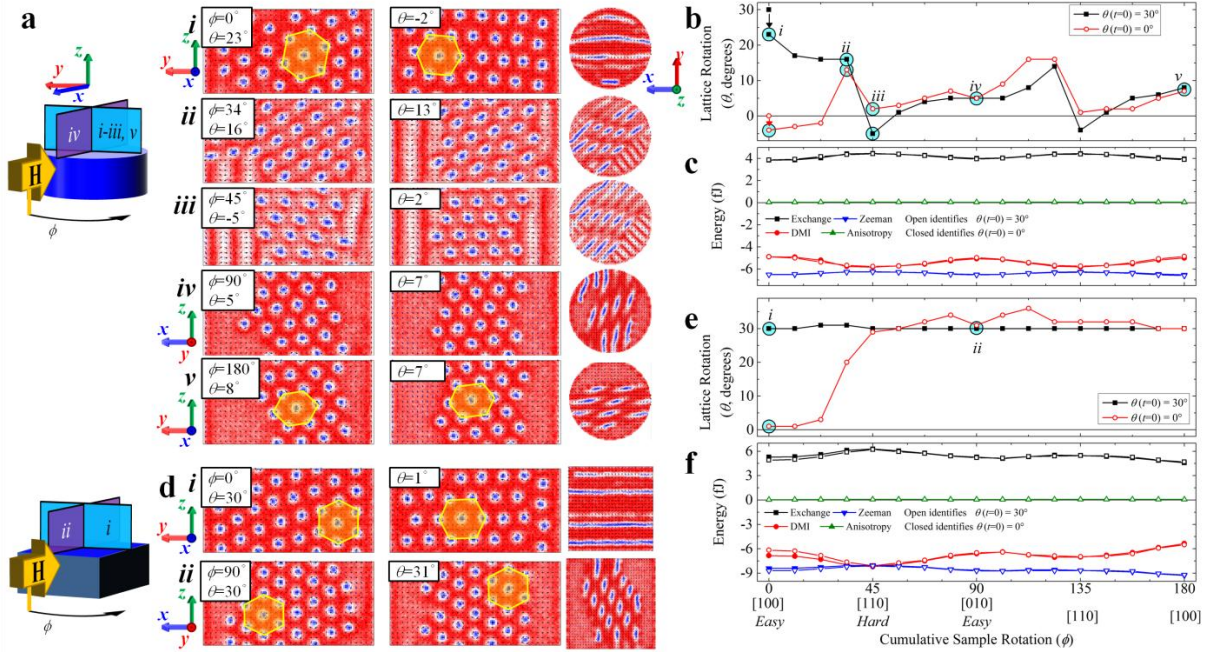

**Figure 6.** OOMMF simulation results: (a) magnetic configuration as seen at the bisection of the cylindrical grain, under increasing $\phi$ (*i-v*), with red indicating the magnetization into the page, and blue indicating out-of-the page. Left and center columns show the skyrmion lattice with initial orientations ($t=0$) of $\theta =30°$ and $0°$, respectively. The lattice is shown from the *x* axis ($\phi=0°$-$45°$, $\phi=135°$-$180°$) or *y* axis ($\phi=56°$-$124°$) to minimize parallax, as shown in the illustrative diagrams; images are 306 nm × 150 nm. Right column shows the magnetization viewed from the direction orthogonal to the plane of rotation; images are 306 nm × 306 nm. (b) Orientation $\theta$ of the skyrmion lattice extracted from the magnetization plots and (c) magnetic energies versus $\phi$. Similarly, (d) the magnetic configuration and (e) skyrmion lattice orientation and (f) magnetic energies are plotted versus $\phi$ for a parallelepiped-shaped grain.

Next, we consider the orientation of the skyrmion lattice during the rotation. The magnetic states at select field angles are shown in Figure 6a, panels *ii – v*, for skyrmion lattices (1) and (2), and the orientation angle of the skyrmion lattices is plotted as a function of sample rotation angle $\phi$ in Figure 6b. These simulation images are from the perspective of the Cartesian axis nearest to the magnetic field, as shown in the illustrative diagrams and in subsequent discussion $\theta$ designates the orientation of the skyrmion axis nearest to the horizontal axis perpendicular to the field, e.g. the $+q_x$ direction. In addition to the identified nearest neighbors there are also nearest neighbors at ±60°. At $\phi=0°$ (panel *i*) the two skyrmion lattices in the left and center columns are rotated by approximately 30° relative to each other, as defined by the initial conditions. The difference in the lattice orientation persists until $\phi=34°$ (panel *ii*), at which point lattice (2) undergoes a dramatic shift to $\theta\approx15°$ to achieve a similar alignment as lattice (1). After rotating the sample to $\phi=45°$ (panel *iii*), corresponding to the [110] magnetic axis, both lattices reorient to $\theta\approx0°$, the initial orientation of lattice (2), with the nearest neighbors in the plane of the rotation (While the [110] direction is not the magnetic hard axis, it is the axis closest to the magnetic hard axis in this plane of rotation). This orientation is directly analogous that observed for the $Cu_2OSeO_3$ samples in Figure 5. Rotating beyond $\phi=45°$ (panel *iv-vi*) the lattices retain this approximate orientation (to within 7°). Enhanced ordering of the skyrmion lattice around $\phi=45°$ rotation is thus consistent with the experimental results for the (Fe, Co)Si sample in Figure 2a, which shows a pronounced reduction in the azimuthal peak width of the skyrmions at this angle, and little change to the skyrmion lattice for $\phi>45°$. In the simulations a smaller restructuring also occurs at $\phi=135°$, again passing through the [110] magnetic hard axis, with both lattices retaining a common orientation and returning to approximately $\theta=0°$ for $\phi\geq135°$. Throughout the rotation there are helical domains which form with periodicity parallel to the magnetic field, particularly on the edges of the simulation volume where the cylindrical curvature is too large to support skyrmions. These helical domains would be invisible in the SANS measurements with the neutrons parallel to the magnetic field, but would appear along the magnetic field direction when viewed (e.g. the neutron path is) orthogonal to the magnetic field. Indeed, Figure 1c and Figure S2, which show the SANS pattern measured orthogonal to the magnetic field, have features located along the magnetic field direction, $q_x$ identifying helical domains.

The simulation results suggest that near the magnetic easy axis the skyrmion lattice prefers a lattice orientation that has the nearest neighbors adjacent in the plane of rotation ($\theta\leq\pm5°$), as observed for the powdered $Cu_2OSeO_3$ sample. In this orientation the skyrmions push directly against their neighbors during the rotation due to the repulsive skyrmion-skyrmion interaction.[48] Starting along a magnetic easy axis, a dramatic reorientation of the skyrmion lattice occurs as the field direction approaches 45°, which corresponds to a magnetic hard axis, although the same reorientation occurs even in the case of relatively weak, or even zero, $K_U$. The energies within the system, shown in Figure 6c, correspondingly show inflections at 45° for both initial lattice orientations. Along this direction the exchange energy increases while the Zeeman decreases and the DMI grows larger (more negative), indicating an enhanced divergence away from the magnetic field direction, and notably, the anisotropy energy remains constant at approximately zero. Of particular note, there is little difference in the energies for the two models with different initial orientations at $\phi=0$ (and throughout the entire $\phi$ range), indicating that the motivating energetics for generating a preferred lattice orientation is dynamic and is not present after the lattice relaxes with the skyrmion tubes aligned with the magnetic field. This behavior is consistent with a driving force of repulsive skyrmion-skyrmion interactions, which would manifest strongly during the dynamic rotation, but would be reduced in the steady-state configuration. Since a reorientation occurs every time the system passes through a magnetic hard axis and the orientation afterwards is the same, rotating the sample repeatedly through hard axes may promote order even in highly-jammed systems. Indeed, this is observed in (Fe, Co)Si which continued to show improved ordering after passing through the [110] direction a second time (between $\phi=90°$ and 180°) and in the $Cu_2OSeO_3$, which required repeated treatments for a total rotation of $\phi=1440°$ (Figure 5). Another possible explanation is that forcing the moments to lie along the magnetic hard axis establishes an effective potential energy. After passing the magnetic hard axis, the orientation with nearest neighbors adjacent in the plane of rotation may be the most effective in reducing this energy. Simulations were also performed (not shown) starting along the magnetic hard axis, analogous to Figure 4, and show a relaxation of the lattice to the same orientation as Figure 6a-*v*. However, for the hard axis simulations, the orientation occurs smoothly and achieves near-full orientation by $\phi=34°$, approaching the magnetic easy axis. These results, coupled with the skyrmion reorientation at simulated $K_u = 0$, suggest there is an additional contribution potentially attributable to skyrmion-skyrmion interactions.

Simulations were also performed for a rectangular parallelepiped-shaped grain under a rotating magnetic field, Figure 6d. This grain will possess additional shape anisotropies absent in the cylindrical grain. As in Figure 6a, the system was initially configured with the magnetization parallel to the magnetic easy axis, with $\theta(t=0)=0°$ and $30°$ in the left and center column, respectively. Once-more the skyrmion tubes follow the rotating magnetic field and undergo a reorientation at $\phi=45°$. However, for the parallelepiped-shaped grain, the reoriented lattice has the nearest neighbors aligned along the axis of rotation, Figure 6d-*ii*, a 30° rotation relative to Figure 6a-*v*. This lattice orientation is directly analogous to the experimental results for the (Fe, Co)Si sample reported in Figure 1 and Figure 4. The collated lattice orientation, Figure 6e, confirms both lattices re-orient to $\theta=30°$ and retain this orientation upon further rotation. Once more, the 30° rotation cannot be explained by previous work.[40] The magnetic energies for the parallelepiped-shaped grain, Figure 6f, have qualitatively similar features as Figure 6c, with inflections around $\phi = 45°$ and again show no significant changes in the anisotropy energies. The difference in the resultant orientation between Figures 6a-*v* and Figure 6d-*ii* could be the result of a larger energy barrier at $\phi=45°$ from the geometry of the simulated grain. However, simulations with large cubic anisotropy ($K_U=500\times10^3$ J/m$^3$) performed on a cylindrical grain show an orientation following Figure 6a, suggesting this is not the case. As another potential source, the cubic simulations show that the skyrmions develop a kink near the grain surface when approaching $\phi=45°$, presumably to remain orthogonal to the surface. This kink may break the symmetry of the skyrmion lattice and give rise to the preferred orientation with $\theta=30°$.

Of further interest is the consideration of topology in these systems. Specifically, initial rotations of the (Fe, Co)Si crystal preserve the ring intensity and decrease the azimuthal width of the six-fold feature, essentially promoting a common orientation of the existing skyrmion arrays. This transformation is primarily an in-plane rotation of the skyrmion axis which preserves topological charge. Additional rotations reduce the ring feature and increase the six-fold pattern, consistent with the breakup of labyrinth domains and formation of skyrmions. This process does not preserve topological charge, as the number of topological structures before and after the rotation are different. Previous works have identified the mechanism for breaking labyrinth domains into skyrmions via nucleation of zero-moment magnetic monopoles.[12] Monopoles are likely nucleated in the plane of the labyrinth domains and propagate along the magnetic field direction, splitting each domain into two smaller bubble domains. Indeed, these same zero-moment monopoles are observed in our simulations. In addition, the destruction of the skyrmions is realized through the nucleation of a magnetic monopole at the end of the skyrmion tube. In the simulations the skyrmions dynamically annihilate via two mechanisms. In the first mechanism a skyrmion breaks in two along its length, and at the bottom of each tube is a Bloch-point monopole, with equal and opposite charge. The monopole propagates along the skyrmion until it reaches the surface and annihilates. In the second mechanism, the skyrmion detaches from the surface of the grain, once more nucleating a Bloch-point monopole, which propagates along the skyrmion tube and annihilates at the opposite surface. Thus, beyond the intriguing symmetry breaking and topological charge considerations, the dynamic break-up of the labyrinth domain and annihilation of the skyrmions is accompanied by the nucleation of charged and un-charged magnetic monopoles.

In summary, we demonstrate the ability to precipitate ordered, oriented skyrmion lattices from otherwise a disordered chiral system consisting of rotationally disordered skyrmion lattices, labyrinth domains and randomly-oriented helices by rotating the sample in a static magnetic field. Rotating a single-crystal ingot of (Fe, Co)Si showed a combination of topological charge-conserving lattice reorientation, and skyrmion formation which constitutes a change in the net topological charge of the system. Interestingly, a similar enhancement of the ordering and collective orientation was observed starting along the high-symmetry [100] easy-axis or the [110] low-symmetry hard axis. This insensitivity to initial orientation was taken to the limit of complete orientation disorder in a polycrystalline sample of $Cu_2OSeO_3$. Rotating the powder sample also precipitated ordered, oriented skyrmion lattices. Micromagnetic simulations confirmed the enhanced orientation and promotion of a preferred orientation for different initial orientations with the most pronounced changes occurring when the field is swept through the magnetic hard axis. In all three cases, the dynamic transformation from labyrinth domains to skyrmion lattices constitutes a change in the topological charge of the system by increasing the number of closed topological structures. The ability to precipitate ordered, oriented skyrmion lattices in highly-disordered systems enables the study of skyrmion states even in polycrystalline or powdered samples, greatly improving the ability to rapidly screen new candidate skyrmion materials.

**Acknowledgements**

D.A.G. acknowledge support from the NRC RAC program, and DOC. We appreciate instrument support from Jeff Kryzwon, Markus Bleuel and Tanya Dax. This material is based upon activities supported by the National Science Foundation under Agreement No. DMR-9986442. P.N. was sponsored by the Center for High Resolution Neutron Scattering as part of the NIST Summer Research Fellowship program, No. NSF DMR 1508249.

# Supplemental Material for:

# Precipitating Ordered Skyrmion Lattices from Helical Spaghetti

Dustin A. Gilbert,[1,2 *] Alexander J. Grutter,[1] Paul M. Neves,[3] Guo-Jiun Shu,[4,5] Gergely Zimanyi,[6]

Brian B. Maranville,[1] Fang-Cheng Chou,[4] Kathryn Krycka,[1] Nicholas P. Butch,[1,3]

Sunxiang Huang,[6] Julie A. Borchers[1]

[1] *NIST Center for Neutron Research, National Institute of Standards and Technology, Gaithersburg, MD, 20899*
[2] *Department of Materials Science and Engineering, University of Tennessee, Knoxville, TN*
[3] *Center for Nanophysics and Advanced Materials, Department of Physics, University of Maryland, MD*
[3] *Center for Condensed Matter Sciences, National Taiwan University, Taipei, 10617 Taiwan*
[4] *Department of Materials and Mineral Resources Engineering, Institute of Mineral Resources Engineering, National Taipei University of Technology, Taipei 10608, Taiwan*
[5] *Department of Physics, University of California, Davis, CA*
[6] *Department of Physics, University of Miami, FL*

*dagilbert@utk.edu

## Examples SANS patterns from idealized systems

In the main text we identify a six-fold feature which many previous works have shown to be associated with the skyrmion ordering, with the orientation of the 6-fold pattern attributable to the orientation of the skyrmion array. Furthermore, we identify a ring feature which may be generated by some combination of labyrinth domain states, randomly oriented helixes, or rotationally disordered skyrmion states. In Figure S1 we simulate these magnetic structures using OOMMF, then Fourier transformed the simulated pattern using ImageJ. The Fourier transform pattern is directly analogous to the SANS pattern. The helical state produces two peaks in the Fourier transform oriented along the axis of the helical propagation, while a multidomain state of misoriented helices with the same periodicity manifests a series of peaks forming a ring. The labyrinth domain, which can be considered as the small-domain limit of a randomly oriented helical state, forms a ring as well, consistent with the claims in the main text. Panels (d) and (e) show the skyrmion structure and confirm that the Fourier transform is composed of six hexagonally-ordered peaks. Furthermore, the lattice orientations are reflected in the corresponding orientations of the Fourier transform. The Fourier transforms of the idealized skyrmion state manifest as sharp peaks, panels (f) and (g) illustrate the effect of introducing a distribution of skyrmion periodicities and rotations respectively. Introducing a 15% distribution of periodicities acts to distort the Fourier transform in the radial direction, while introducing a ±12° rotational distribution to the lattice distorts the Fourier transform in the azimuthal direction. Similar to panel (b), panel (h) is a multidomain state consisting of the rotationally distributed skyrmion state from (g) and the labyrinth state in (c). The calculated Fourier transform very accurately reproduces our experimentally observed SANS pattern.

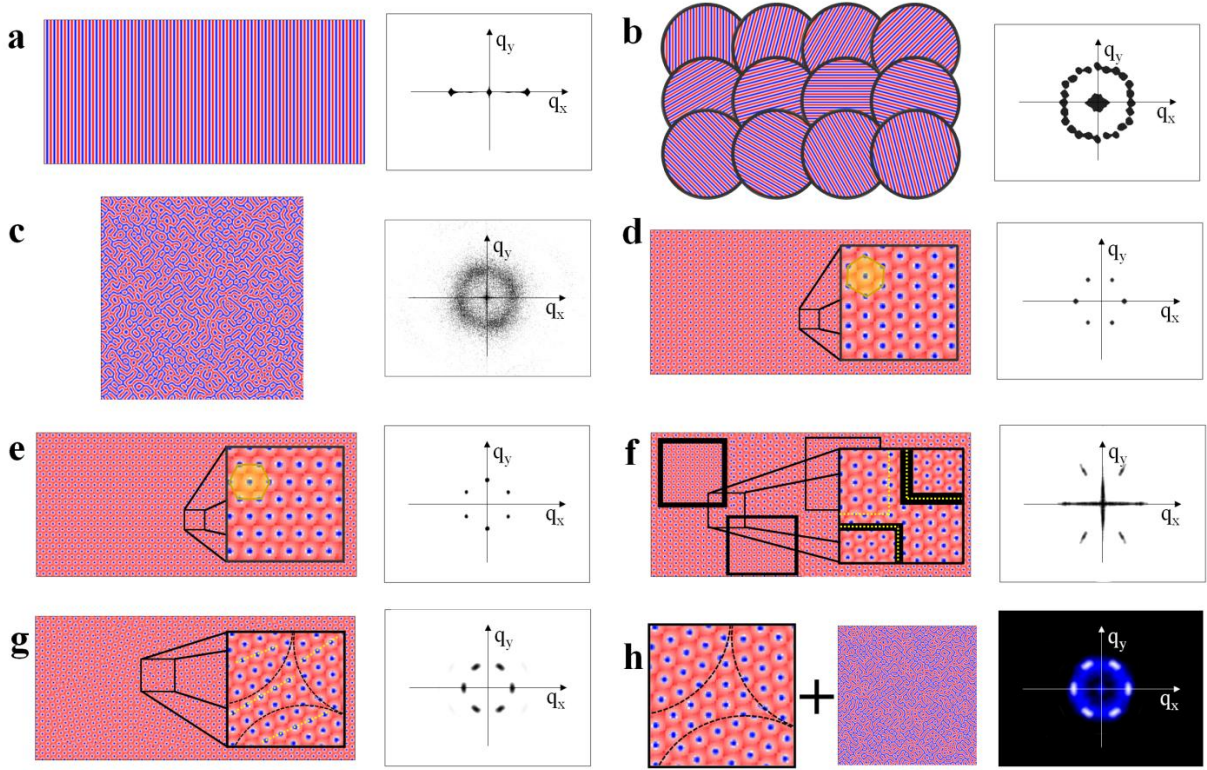

**Figure S1.** Simulated magnetic profiles and Fourier transform of patterns of (a) helical state, (b) multidomain helical state, (c) labyrinth domain state, (d) skyrmion domain with nearest neighbors vertically aligned, (e) skyrmion domain with nearest neighbors laterally aligned, (f) multidomain skyrmion configuration with 15% distribution in the periodicity and (g) ±12° distribution in the rotation of the skyrmion lattice, and (h) modeled multidomain consisting of (c) and (g).

**SANS orthogonal to the skyrmion array**

In these measurements the SANS technique is sensitive to magnetic structures which are periodic in the plane orthogonal to the neutron beam propagation direction. In the main text, we focus on measurements performed with the neutron beam parallel to the magnetic field. In this configuration the skyrmion tubes form parallel to the neutron beam, arranged in hexagonal arrays which are orthogonal to the neutron beam. As it is known that the jammed magnetic state is quite complex, we also perform measurements on the (Fe, Co)Si crystal with the neutron beam orthogonal to the magnetic field and therefore the skyrmion tubes, Figure S2. The magnetic state of these measurements was prepared in the same way as Figure 1 in the main text, at 17.5 K, 40 mT, after zero-field cooling and saturation. Measurements performed in this orientation show two peaks along $q_x$ and two along $q_y$. The peaks along $q_x$ indicate a periodicity along the direction of the skyrmion tubes, potentially from a helically structured minority phase. The two peaks aligned along $q_y$ are from the ring structure. The peaks along $q_y$ decrease by 30%, qualitatively similar to the ring structure in the initial orientation. In this geometry the neutrons are sensitive to different orientations of the spins compared to the orientation in Figure 1a, and thus we do not expect the changes to be quantitatively identical.

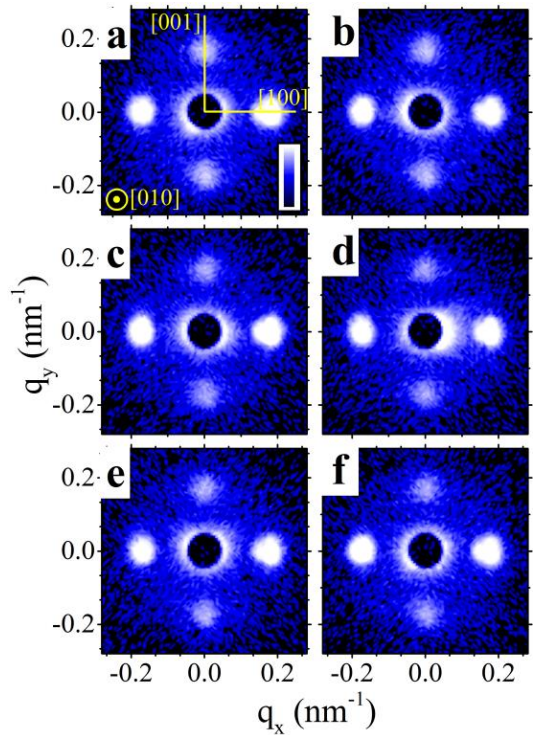

**Figure S2.** SANS images of (Fe, Co)Si taken with the neutron beam perpendicular to the magnetic field, analogous to Figure 1 in the main text, at $\phi =$ (a) 0°, (b) 11.25° (c) 22.5°, (d) 45° (e) 90° and (f) 180°.

## Azimuthal projections of Figure 4

Azimuthal projections were performed on the SANS results for (Fe, Co)Si presented in the main text, Figure 4. A model was fitted using six coupled Gaussian functions, as discussed in the main text, and the integrated area presented as a simple phase fraction, shown in Figure 4g. The fitted profiles are shown here, in Figure S3.

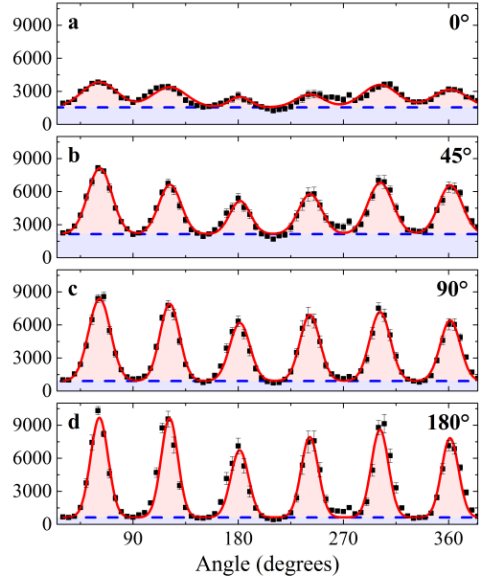

**Figure S3** Azimuthal projections of the data from the main text, Figure 4. Measurements were performed at $\phi =$ (a) 0°, (b) 45°, (c) 90°, (d) 180°

**Rotation performed at lower field**

In the main text, the skyrmion rotation of (Fe, Co)Si was performed at 17.5 K in a static magnetic field of 40 mT, after zero-field cooling, then saturating the sample at 500 mT. Similar measurements were performed by preparing the skyrmion state by zero-field cooling to 20 K, applying a 500 mT saturating field, then reducing the field to 25 mT. As in Figure 1 of the main text, the magnetic field and neutron beam are initially parallel to the crystalline [100] direction. In this configuration the SANS pattern shows a ring structure with a weak 6-fold skyrmion pattern, similar to Figure 1b in the main text. As in the main text, rotating the sample in the static magnetic field strengthens the 6-fold pattern relative to the ring feature, Figure S4a-c. The azimuthal projection of the SANS patterns, Figure S4d, shows that rotating the sample clearly decreases the strength of the ring feature. Fitting the data using the coupled Gaussian function defined in the main text shows that the area of the skyrmion feature increases by 80% over the 180° $\phi$ rotation; the skyrmion phase fraction increases from 11% in the as-prepared state to 38% at $\phi = 180°$. This indicates that the ordering sequence is effective throughout the skyrmion stability envelope. It is unclear at this point if these results indicate improved efficacy of the technique at higher fields and lower temperatures, or if the reduced efficacy is due to these measurements being performed on the boundary of the stability envelope.

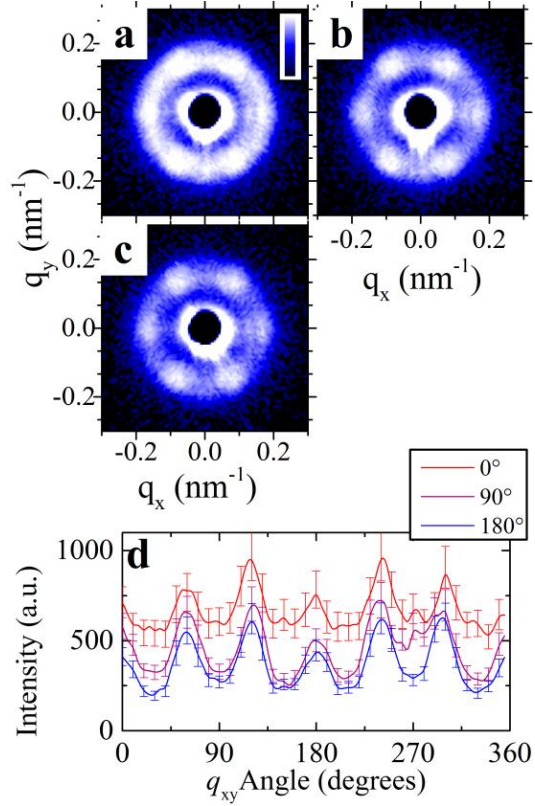

**Figure S4** SANS images of (Fe, Co)Si taken at and 20 K, 25 mT, and $\phi =$ (a) 0°, (b) 90°, and (c) 180°. (d) Azimuthal projections of the SANS images in panels (a) - (c).

**Rotation in the pseudo stable window**

The ability to realize skyrmions outside of their traditional stability envelope was investigated by performing the skyrmion rotation at 12 K and 40 mT. First, the sample was heated to 17.5 K, then saturated, then the field was reduced to 40 mT, following our initial procedure. The sample was then field cooled to 12 K in 40 mT. The resultant feature, in Fig. S5a, shows a ring feature with subtle six-fold symmetry, consistent with the structure discussed in the main text. The sample was then rotated through $\phi = 900°$, Fig. S5b, resulting in a weaker ring, with higher intensities along the axis of rotation. This likely indicates that some of the moments form a conical phase along the neutron path, as would be the case coming from zero-field cooling or saturation, with some residual domains, particularly aligned along the axis of rotation. The residual moments along this axis are somewhat expected as the sample rotation is energetically degenerate along this axis. In a second test, the sample was saturated at 17.5 K, then the field was reduced to 40 mT, then the sample was rotated through $\phi = 360°$, resulting in the ordered phase discussed in the main text. Field cooling the sample to 12 K the sample retained the ordered skyrmion phase as confirmed in Fig. S5c. Rotating the sample at 12 K in this configuration, Fig. S5d, resulted again in a weak ring with higher intensities at the top and bottom, as before. Lastly, the sample was saturated at 12 K in a 300 mT field along the neutron path, then the field was reduced to 40 mT, forming a conical state along the neutron path, which showed very weak signal with no structure. The sample was then rotated through $\phi = 180°$, Fig. S5e, and $\phi = 900°$, Fig. S5f. The resultant structure showed a weak ring with intensities along the top and bottom. The intensity of the feature seems to increase with iterative rotation, indicating an enhancement in the disorder potentially due to pinning of the conical phase as it is rotated through the sample.

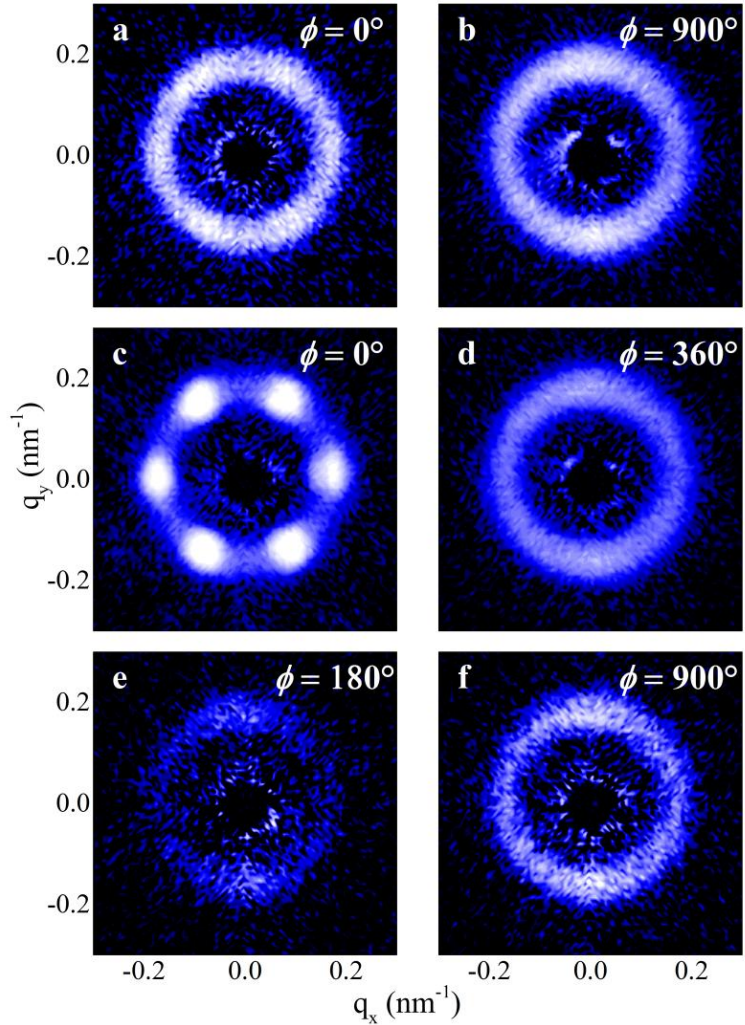

**Figure S5** SANS images of (Fe, Co)Si rotated in the pseudo stable skyrmion window ($T=12$ K, $\mu_0 H = 40$ mT) (a) field cooling from $T > T_C$, and (b) after $\phi = 900°$ rotation; (c) after preparing the skyrmion state at $T=17.5$ K and $\mu_0 H = 40$ mT by 360° rotation, then field cooling to $T=12$ K, $\mu_0 H = 40$ mT and (d) rotating $\phi = 360°$; after applying a saturating field (300 mT) along the neutron direction at 12 K, then (e) rotating $\phi = 180°$ and (f) 900°. Intensity scale in left and right column are set at the max intensity and a common minimum (20 counts) for each row.

**Rotation of MnSi Single Crystal**

A large single crystal MnSi was rotated in the (001) plane in supplemental experiments, shown in Figure S6. Similar to the materials featured in the main text ((Fe, Co)Si and Cu$_2$OSeO$_3$), saturating MnSi and returning to 180 mT results in a ring-like feature. Rotating the sample causes a transition between six-fold and 12-fold patterns. These patterns indicate the sample possesses one and two dominant orientations of the skyrmion domains, following the magnetocrystalline easy axes. This is in agreement with the work of Luo et al., cited in the main text.

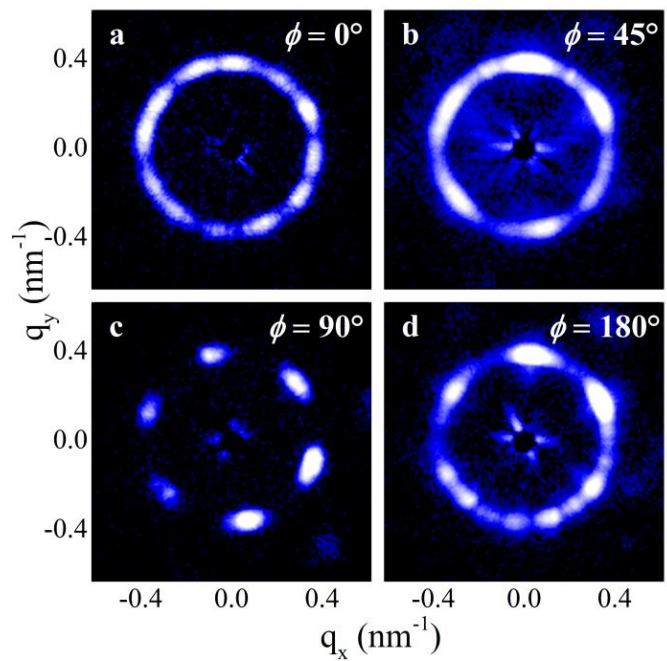

**Figure S6** SANS images of MnSi taken at 29 K and 180 mT, and $\phi$ = (a) 0°, (b) 45°, (c) 90° and (d) 180°, starting aligned with the [100] direction.